\documentstyle[prc,aps,preprint]{revtex}
\begin{document}
\draft
\date{\today}
\title{The decay $\rho^{0}\rightarrow \pi^{+}+\pi^{-}+\gamma$ and the coupling constant
g$_{\rho\sigma\gamma}$ }

\author{A. Gokalp~\thanks{agokalp@metu.edu.tr} and
        O. Yilmaz~\thanks{oyilmaz@metu.edu.tr}}
\address{ {\it Physics Department, Middle East Technical University,
06531 Ankara, Turkey}}
\maketitle

\begin{abstract}
The experimental branching ratio for the radiative decay
$\rho^{0}\rightarrow\pi^{+}+\pi^{-}+\gamma$ is used to estimate the coupling constant
$g_{\rho\sigma\gamma}$ for a set of values of $\sigma$-meson parameters M$_{\sigma}$
and $\Gamma_{\sigma}$. Our results are quite different than the values of this constant used
in the literature.
\end{abstract}

\thispagestyle{empty}
~~~~\\
%\narrowtext
\newpage
\setcounter{page}{1}
%%%
%%%

The radiative decay process $\rho^{0}\rightarrow \pi^{+}+\pi^{-}+\gamma$
has been studied employing different approaches \cite{R1,R5}. There are
two mechanisms that can contribute to this radiative decay: the first one
is the internal bremsstrahlung where one of the charged pions from the decay
$\rho^{0}\rightarrow \pi^{+}+\pi^{-}$ emits a photon, and the second one is
the structural radiation which is caused by the internal transformation of
the $\rho$-meson quark structure. Since the bremsstrahlung is well described
by quantum electrodynamics, different methods have been used to estimate
the contribution of the structural radiation.

Singer \cite{R1} calculated the amplitude for this decay by considering only
the bremsstrahlung mechanism since the decay
$\rho^{0}\rightarrow \pi^{+}+\pi^{-}$ is the main decay mode of
$\rho^{0}$-meson. He also used the universality of the coupling of the
$\rho$-meson to pions and nucleons to determine the coupling constant
$g_{\rho\pi\pi}$ from the knowledge of the coupling constant
$g_{\rho NN}$. Later, Renard \cite{R3} studied this decay among
other vector meson decays into $2\pi+\gamma$ final states in a gauge
invariant way with current algebra, hard-pion and Ward-identities techniques.
He, moreover, established the correspondence between these current algebra
results and the structure of the amplitude calculated in the
single particle approximation for the intermediate states. In corresponding
Feynman diagrams the structural radiation proceeds through the
intermediate states as $\rho^{0}\rightarrow S+\gamma$ where the meson S
subsequently decays into a $\pi^{+}\pi^{-}$ pair. He concluded that
the leading term is the pion bremsstrahlung and that the largest
contribution to the structural radiation amplitude results from the
scalar $\sigma$-meson intermediate state. He used the rough estimate
$g_{\rho\sigma\gamma}\simeq 1$ for the coupling constant
$g_{\rho\sigma\gamma}$ which was obtained with the spin independence
assumption in the quark model. The coupling constant $g_{\rho\pi\pi}$
was determined using the then available experimental decay rate of
$\rho$-meson and also current algebra results as
$3.2\leq g_{\rho\pi\pi}\leq 4.9$. On the other hand, the coupling constant
$ g_{\sigma\pi\pi}$ was deduced from the assumed decay rate
$\Gamma\simeq 100$ MeV for the $\sigma$-meson as $ g_{\sigma\pi\pi}$=3.4
with $M_{\sigma}=400$ MeV. Furthermore, he observed that the $\sigma$-
contribution modifies the shape of the photon spectrum for high momenta
differently depending on the mass of the $\sigma$-meson.
We like to note, however, that the nature of the $\sigma$-meson
as a $\bar{q} q$ state in the naive quark model and therefore the estimation
of the coupling constant $g_{\rho\sigma\gamma}$ in the quark model have been
a subject of controversy. Indeed, Jaffe \cite{R6,R7} lately argued within
the framework of lattice QCD calculation of pseudoscalar meson scattering
amplitudes that the light scalar mesons are $\bar{q}^{2}q^{2}$ states rather
than $\bar{q} q$ states.

Recently, on the other hand, the coupling constant $ g_{\rho\sigma\gamma}$
has become an important input for the studies of $\rho^{0}$-meson
photoproduction  on nucleons. The presently available data \cite{R8}
on the photoproduction of $\rho^{0}$-meson on proton targets near threshold
can be described at low momentum transfers by a simple one-meson exchange
model \cite{R9}. Friman and Soyeur \cite{R9} showed that in this picture
the $\rho^{0}$-meson photoproduction cross section on protons is given
mainly by $\sigma$-exchange. They calculated the $\gamma\sigma\rho$-vertex
assuming Vector Dominance of the electromagnetic current, and their result
when derived using an effective Lagrangian for the
$\gamma\sigma\rho$-vertex gives the value $g_{\rho\sigma\gamma}\simeq$2.71
for this coupling constant. Later, Titov et al. \cite{R10} in their study
of the structure of the $\phi$-meson photoproduction amplitude based on
one-meson exchange and Pomeron-exchange mechanisms used the coupling
constant $g_{\phi\sigma\gamma}$ which they calculated from the above value
of $g_{\rho\sigma\gamma}$ invoking unitary symmetry arguments as
$g_{\phi\sigma\gamma}\simeq$0.047. They concluded that the data at low
energies near threshold can accommodate either the second Pomeron or
the scalar mesons exchange, and the differences between these competing
mechanisms have profound effects on the cross sections and the polarization
observables.

It, therefore, appears of much interest to study the coupling constant
$g_{\rho\sigma\gamma}$ that plays an important role in scalar meson
exchange mechanism from a different perspective other than Vector Meson Dominance
as well. For this purpose we calculate the branching ratio for the radiative
decay $\rho^{0}\rightarrow \pi^{+}+\pi^{-}+\gamma$, and using
the experimental value $0.0099\pm 0.0016$ for this branching ratio \cite{R11},
we estimate the coupling constant $g_{\rho\sigma\gamma}$. Our calculation is based
on the Feynman diagrams shown in Fig. 1. The first two terms in this figure
are not gauge invariant and they are supplemented by the direct term shown
in Fig. 1(c) to establish gauge invariance. Guided by Renard's \cite{R3}
current algebra results, we assume that the structural radiation amplitude
is dominated by $\sigma$-meson intermediate state which is depicted in Fig. 1(d).

We describe the $\rho\sigma\gamma$-vertex by the effective Lagrangian
\begin{eqnarray}
{\cal L}^{int.}_{\rho\sigma\gamma}=\frac{e}{M_{\rho}}g_{\rho\sigma\gamma}
   [\partial^{\alpha}\rho^{\beta}\partial_{\alpha}A_{\beta}
   -\partial^{\alpha}\rho^{\beta}\partial_{\beta}A_{\alpha}]\sigma
\end{eqnarray}
which also defines the coupling constant $g_{\rho\sigma\gamma}$.
The $\rho\pi\pi$-vertex is described by the effective Lagrangian
\begin{eqnarray}
{\cal L}^{int}_{\rho\pi\pi}=
g_{\rho\pi\pi}\vec{\rho}_{\mu}\cdot (\partial^{\mu}\vec{\pi}\times\vec{\pi})
\end{eqnarray}
using which we obtain the decay width $\Gamma(\rho\rightarrow\pi\pi)$ as
\begin{eqnarray}
\Gamma(\rho\rightarrow\pi\pi)=\frac{g^{2}_{\rho\pi\pi}}{4\pi}\frac{M_{\rho}}{12}
\left [ 1-(\frac{2M_{\pi}}{M_{\rho}})^{2}\right ] ^{3/2}~~.
\end{eqnarray}
The experimental value of the width $\Gamma$=151 MeV \cite{R11} then yields
the value $\frac{g^{2}_{\rho\pi\pi}}{4\pi}$=2.91 for the coupling constant
$g_{\rho\pi\pi}$. For the $\sigma\pi\pi$-vertex we use the effective
Lagrangian \cite{R12}
\begin{eqnarray}
{\cal L}^{int}_{\sigma\pi\pi}=
\frac{1}{2}g_{\sigma\pi\pi}M_{\sigma}\vec{\pi}\cdot\vec{\pi}\sigma~~.
\end{eqnarray}
The decay width of the $\sigma$-meson that follows from this effective
Lagrangian is given as
\begin{eqnarray}
\Gamma_{\sigma}\equiv\Gamma(\sigma\rightarrow\pi\pi)=
\frac{g^{2}_{\sigma\pi\pi}}{4\pi}\frac{3M_{\sigma}}{8}
\left [ 1-(\frac{2M_{\pi}}{M_{\sigma}})^{2}\right ] ^{1/2}~~.
\end{eqnarray}

In our calculation of the invariant amplitude for the process
$\rho^{0}\rightarrow \pi^{+}+\pi^{-}+\gamma$, in the $\sigma$-meson
propagator in Fig. 1(d) we make the replacement
$M_{\sigma}\rightarrow M_{\sigma}-\frac{1}{2}i\Gamma_{\sigma}$,
where $\Gamma_{\sigma}$ is given by Eq. (5). Since the experimental
candidate for $\sigma$-meson $f_{0}$(400-1200) has
a width (600-1000) MeV \cite{R11}, we obtain a set of values for the
coupling constant $g_{\rho\sigma\gamma}$ by considering the ranges
$M_{\sigma}$=400-1200 MeV, $\Gamma_{\sigma}$=600-1000 MeV for the
parameters of the $\sigma$-meson.

In terms of the invariant amplitude ${\cal M}$(E$_{\gamma}$, E$_{1}$),
the differential decay probability for an unpolarized $\rho^{0}$-meson
at rest is given by
\begin{eqnarray}
\frac{d\Gamma}{dE_{\gamma}dE_{1}}=\frac{1}{(2\pi)^{3}}\frac{1}{8M_{\rho}}
\mid {\cal M}\mid^{2} ,
\end{eqnarray}
where E$_{\gamma}$ and E$_{1}$ are the photon and pion energies respectively,
and we perform an average over the spin states of $\rho^{0}$-meson
and a sum over the polarization states of the photon.
The decay width $\Gamma(\rho\rightarrow\pi^{+}+\pi^{-}+\gamma)$ is the
obtained by integration
\begin{eqnarray}
\Gamma=\int_{E_{\gamma,min.}}^{E_{\gamma,max.}}dE_{\gamma}
       \int_{E_{1,min.}}^{E_{1,max.}}dE_{1}\frac{d\Gamma}{dE_{\gamma}dE_{1}}~~~~.
\end{eqnarray}
The maximum photon energy E$_{\gamma, max.}$ is given as
$E_{\gamma,max.}=(M_{\rho}^{2}-4M_{\pi}^{2})/2M_{\rho}$=334 MeV,
and the minimum photon energy is taken as E$_{\gamma,min.}$=50 MeV,
since the experimental value of the branching ratio is determined for
this range of photon energies \cite{R13}. The maximum and minimum values
for pion energy E$_{1}$ are given by
\begin{eqnarray}
\frac{1}{2(2E_{\gamma}M_{\rho}-M_{\rho}^{2})}
[-2E_{\gamma}^{2}M_{\rho}+3E_{\gamma}M_{\rho}^{2}-M_{\rho}^{3}
~~~~~~~~~~~~~~~~~~~~~~~~~~~~\nonumber \\
\pm  E_{\gamma}\sqrt{(-2E_{\gamma}M_{\rho}+M_{\rho}^{2})
       (-2E_{\gamma}M_{\rho}+M_{\rho}^{2}-4M_{\pi}^{2})}~]~.
\nonumber
\end{eqnarray}

The photon spectra for the branching ration of the decay
$\rho^{0}\rightarrow \pi^{+}+\pi^{-}+\gamma$ are plotted in Fig. 2
as a function of photon energy E$_{\gamma}$.
The contributions of bremsstrahlung and structural radiation amplitude
calculated with $\sigma$-meson intermediate state as well as the
contribution of the interference term are shown together with
the phase space for this decay as a function of photon energy.
The parameters of the $\sigma$-meson adapted
for the numerical calculations are M$_{\sigma}$=500 MeV,
$\Gamma_{\sigma}$=800 MeV resulting in the coupling constants
$g_{\sigma\pi\pi}=8.04$ and $g_{\rho\sigma\gamma}=8.45$.
The contribution of the $\sigma$-term becomes increasingly
important in the region of high photon energies dominating the
contribution of the bremsstrahlung amplitude.
Although its contribution is somewhat reduced by the interference term,
the structural radiation makes the main contribution to the branching ratio
in the region of high photon energies. The pion energy spectra for
the branching ratio as a function of pion energy are shown in Fig. 3 for
the same set of $\sigma$-meson parameters. The contribution of the
$\sigma$-meson term is again significant in the region of low pion
energies becoming quite insignificant for high pion energies compared
to bremsstrahlung term and the interference term in general makes
a small but negative contribution.

In Fig. 4 we show the dependence of the branching ratio calculated
with the same parameters for the $\sigma$-meson on the minimum detected photon
energy. This dependence is quite strong and therefore in our calculations we use
minimum detected photon energy as 50 MeV as quoted in the experimental
results \cite{R13}. We furthermore plot the ratio R$_{\beta}$ as a function
of $\beta$ in Fig. 5. This ratio is defined by
\begin{eqnarray}
R_{\beta}=\frac{\Gamma_{\beta}}
{\Gamma_{tot}(\rho^{0}\rightarrow \pi^{+}+\pi^{-}+\gamma)}
\end{eqnarray}
where the numerator and denominator are given as
\begin{eqnarray}
\Gamma_{\beta}=\int_{50}^{50+\beta}dE_{\gamma}\frac{d\Gamma}{dE_{\gamma}}~~, ~~~~
\Gamma_{tot.}=\int_{50}^{E_{\gamma,max.}}dE_{\gamma}\frac{d\Gamma}{dE_{\gamma}}~~.
\end{eqnarray}
The shape of the curve for this ratio further reflects that the
dependence of the branching ratio on the $\sigma$-term is pronounced in the region
of the high photon energies.

We present the results of our calculation in Table 1.
We note that by using the experimental value of the decay rate \cite{R11}
for the radiative decay $\rho^{0}\rightarrow \pi^{+}+\pi^{-}+\gamma$
in the expression we obtain in our calculation for this decay rate, we
arrive at a quadric equation for the coupling constant
$g_{\rho\sigma\gamma}$, the coefficient of the quadric term resulting
from the $\sigma$-meson contribution of Fig. 1(d) and the coefficient
of the linear term from the interference of the $\sigma$-meson and the pion
bremsstrahlung terms of Fig. 1(a,b,c).
Therefore, our analysis produces a set of values for the coupling constant
$g_{\rho\sigma\gamma}$ depending on the $\sigma$-meson parameters,
and for a given set of $\sigma$-meson parameters results in two
values for the coupling constant $g_{\rho\sigma\gamma}$,
one being positive and the other one negative.
Note that from the specific structure of the matrix element for the
$\sigma$-meson contribution we can obtain in a straightforward manner
the approximate relation
\begin{eqnarray}
\frac{g_{\rho\sigma\gamma}^{2}}{\Gamma_{\sigma}M_{\sigma}^{3}}\simeq constant~~~.
\end{eqnarray}
Furthermore, the values of the coupling constant $g_{\rho\sigma\gamma}$ resulting from our
estimation are in general quite different than the values of this constant usually adopted
for the one-meson exchange mechanism calculations existing in the literature.
For example, Titov et al. \cite{R10} uses the value $g_{\rho\sigma\gamma}$=2.71
which they obtain from Friman and Soyeur's \cite{R9} analysis of $\rho$-meson
photoproduction using Vector Meson Dominance. It is interesting
to note that in their study of pion dynamics in Quantum Hadrodynamics II,
which is a renormalizable model constructed using local gauge invariance
based on SU(2) group, that has the same Lagrangian densities for the
vertices we use, Serot and Walecka \cite{R14} come to the conclusion that
in order to be consistent with the experimental result that
s-wave $\pi N$-scattering length is anomalously small, in their tree-level
calculation they have to choose $g_{\sigma\pi\pi}$=12.
Since they use $M_{\sigma}$=520 MeV this implies
$\Gamma_{\sigma}\simeq$ 1700 MeV. If we use these values in our analysis,
we then obtain $g_{\rho\sigma\gamma}$=11.91.
Soyeur \cite{R12}, on the other hand, uses quite arbitrarly
the values $M_{\sigma}$=500 MeV, $\Gamma_{\sigma}$=250 MeV, which
in our calculation results in the coupling constant
$g_{\rho\sigma\gamma}$=6.08.
We like to note, however, that these values for $\sigma$-meson parameters
are not consistent with the experimental data on $\sigma$-meson \cite{R11}.

Our analysis and estimation of the coupling constant $g_{\rho\sigma\gamma}$
using the experimental value of the branching ratio of the radiative decay
$\rho^{0}\rightarrow \pi^{+}+\pi^{-}+\gamma$ give quite different values
for this coupling constant than used in the literature. Furthermore,
since we obtain this coupling constant as a function of $\sigma$-meson
parameters, it will be of interest to study the dependence of the observables
of the reactions, such as for example the photoproduction of vector mesons on
nucleons $\gamma~+~N~\rightarrow N~+~V$ where V is the neutral vector meson,
analyzed using one-meson exchange mechanism on these parameters.

\begin{center}
{\bf Acknowledgments}
\end{center}

We thank  Prof. Dr. M. P. Rekalo for suggesting this problem to us and for
his guidance during the course of our work. We also wish to thank Prof. Dr.
T. M. Aliev for helpful discussions.

%\pagebreak

\begin{table}
\caption{The calculated coupling constant $g_{\rho\sigma\gamma}$
for different $\sigma$-meson parameters}
\begin{tabular}{|c|c|c|c|c|}
$M_{\sigma}$ (MeV)&$\Gamma_{\sigma}$ (MeV)&$g_{\sigma\pi\pi}$
&$g_{\rho\sigma\gamma}$ &$g_{\rho\sigma\gamma}$\\
\hline
500 & 600 & 6.97 & 7.64$\pm$ 1.56 & -6.00$\pm$ 1.58 \\ \hline
500 & 800 & 8.04 & 8.45$\pm$ 1.77 & -6.96$\pm$ 1.78\\ \hline
600 & 600 & 6.16 & 9.83$\pm$ 1.85 & -6.68$\pm$ 1.85\\ \hline
600 & 800 & 7.11 & 10.49$\pm$ 2.07 & -7.70$\pm$ 2.06 \\ \hline
800 & 600 & 5.18 & 15.05$\pm$ 2.64 & -9.11$\pm$ 2.64\\ \hline
800 & 900 & 6.34 & 15.29$\pm$ 2.84 & -10.17$\pm$ 2.84 \\ \hline
900 & 600 & 4.85 & 18.14$\pm$ 3.13 & -10.65$\pm$ 3.14 \\ \hline
900 & 900 & 5.94 & 17.78$\pm$ 3.23 & -11.35$\pm$ 3.23
\end{tabular}
\end{table}

\newpage

{\bf Figure Captions:}

\begin{description}

\item[{\bf Figure 1}:] Diagrams for the decay $\rho^{0}\rightarrow \pi^{+}+\pi^{-}+\gamma$

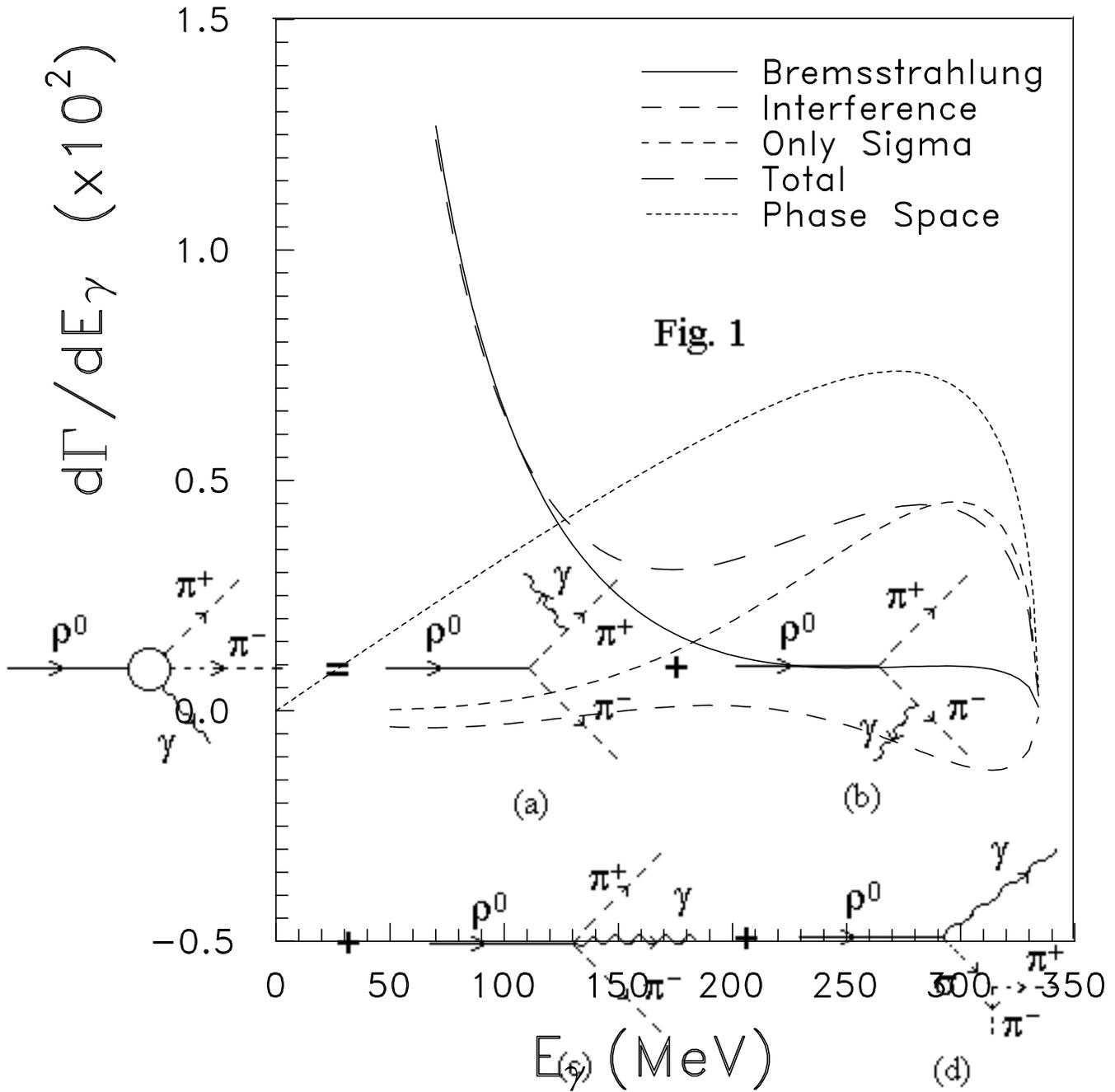
\item[{\bf Figure 2}:] The photon spectra for the decay width of $\rho^{0}\rightarrow
\pi^{+}+\pi^{-}+\gamma$. The contributions of different terms are indicated.

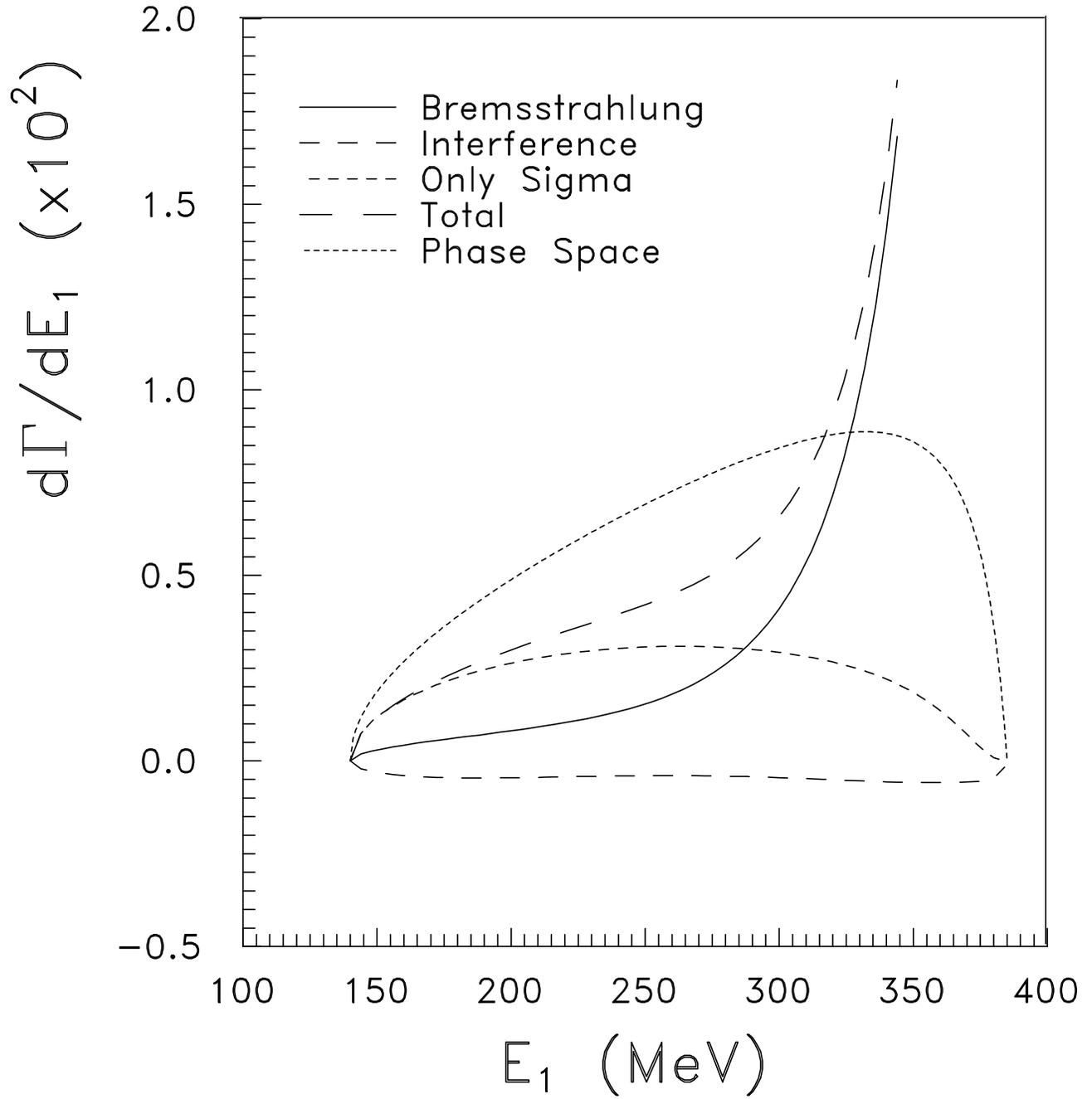
\item[{\bf Figure 3}:] The pion energy spectra for the decay width of
$\rho^{0}\rightarrow\pi^{+}+\pi^{-}+\gamma$.
The contributions of different terms are indicated.

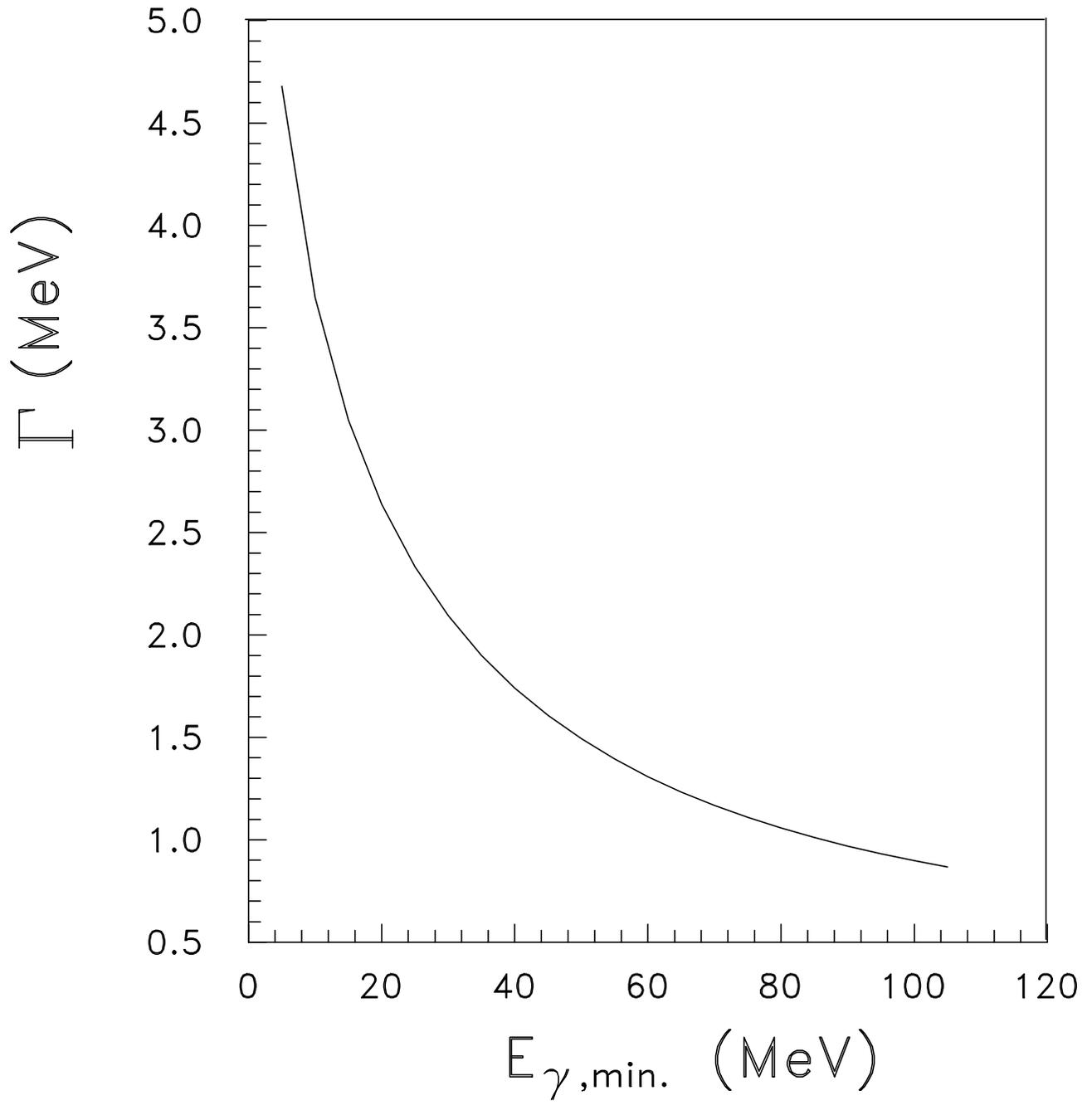
\item[{\bf Figure 4}:] The decay width of $\rho^{0}\rightarrow
\pi^{+}+\pi^{-}+\gamma$ as a function of minimum detected photon energy.

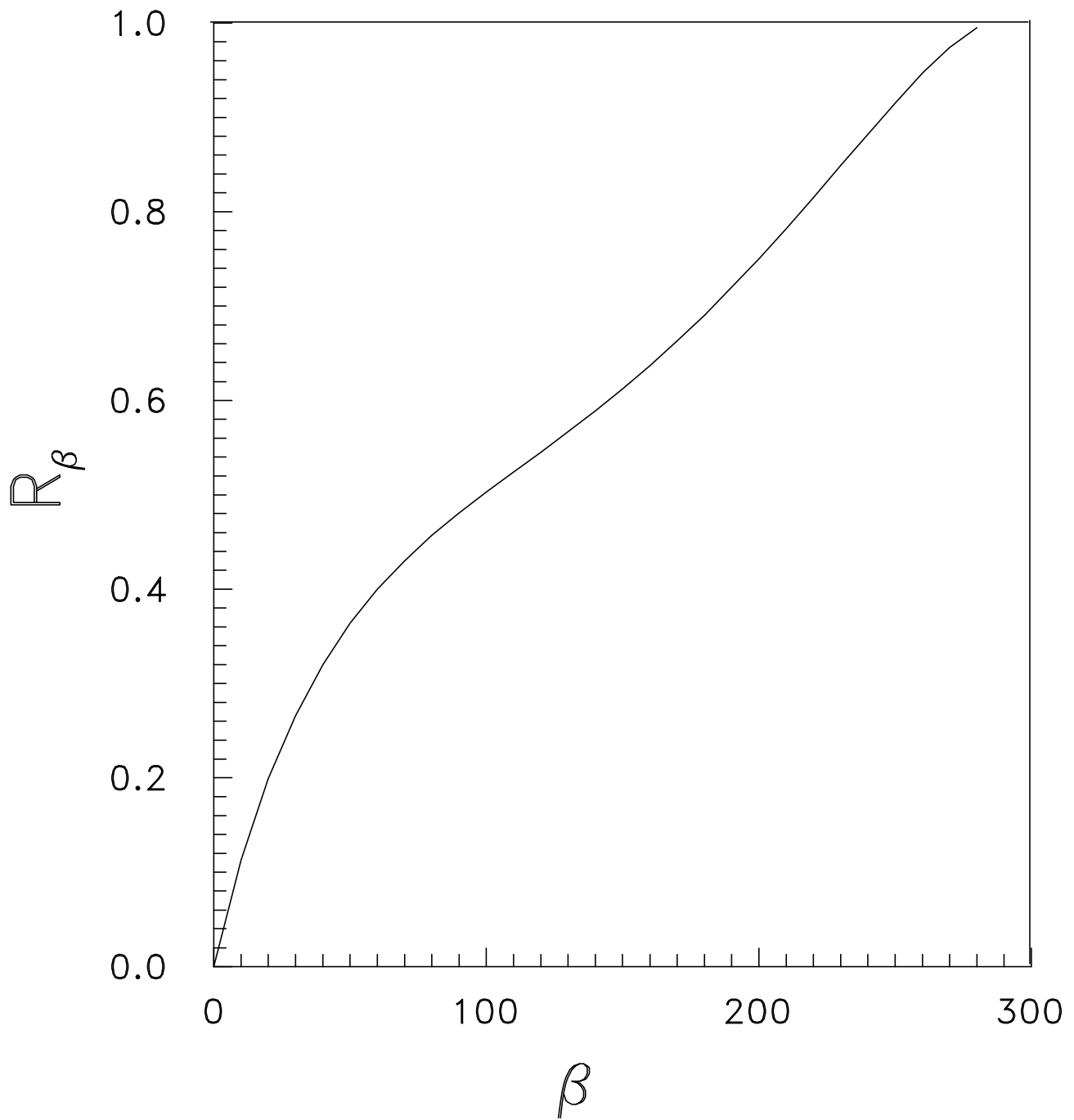
\item[{\bf Figure 5}:] The ratio
R$_{\beta}=\frac{\Gamma_{\beta}}{\Gamma_{tot}}$ as a function of $\beta$.
\end{description}


\begin{references}
\bibitem{R1} P. Singer, Phys. Rev.  {\bf 130} (1963) 2441; {\bf 161} (1967) 1694.
\bibitem{R2} V. N. Baier and V. A. Khoze, Sov. Phys. JETP {\bf 21} (1965) 1145.
\bibitem{R3} S. M. Renard, Nuovo Cim. {\bf 62 A} (1969) 475.
\bibitem{R4} K. Huber and H. Neufeld, Phys. Lett. {\bf B 357} (1995) 221.
\bibitem{R5} E. Marko, S. Hirenzaki, E. Oset and H. Toki, Phys. Lett.
             {\bf B 470} (1999) 20.
\bibitem{R6} R. L. Jaffe, hep-ph/0001123.
\bibitem{R7} M. Alford and R. L. Jaffe, hep-lat/0001023.
\bibitem{R8} Aachen-Berlin-Bonn-Hamburg-Heidelberg-Munchen Collaboration,
            Phys. Rev. {\bf 175} (1968) 1669.
\bibitem{R9} B. Friman and M. Soyeur, Nucl. Phys. {\bf A 600} (1996) 477.
\bibitem{R10} A. I. Titov, T. -S. H. Lee, H. Toki and O. Streltrova, Phys. Rev.{\bf C 60} (1999) 035205.
\bibitem{R11} Review of Particle Physics, Eur. Phys. J.{\bf C 3} (1998) 1.
\bibitem{R12} M. Soyeur, nucl-th/0003047.
\bibitem{R13} S. I. Dolinsky, et al, Phys. Rep.{\bf 202} (1991) 99.
\bibitem{R14} B. D. Serot and J. D. Walecka, in Advances in Nuclear Physics,
             edited by J. W. Negele and E. Vogt,{\bf Vol. 16} (1986).

\end{references}
\end{document}